\documentclass[prb,twocolumn,amsmath,amssymb,floatfix,superscriptaddress]{revtex4-1}
\usepackage{graphicx}
\usepackage[FIGTOPCAP]{subfigure}
\usepackage[usenames]{color}
\usepackage{hyperref}
\usepackage{bbm}

\begin{document}
\title{Band-structure-dependence of renormalization-group prediction on pairing channels}
\date{\today}
\author{Yi-Ting Hsu}
\affiliation{Department of Physics, Cornell University, Ithaca, New York 14853, USA}
\author{Alejandro Federico Rebola}
\affiliation{School of Applied and Engineering Physics, Cornell University, Ithaca, New York 14853, USA}
\author{Craig J. Fennie}
\affiliation{School of Applied and Engineering Physics, Cornell University, Ithaca, New York 14853, USA}
\author{Eun-Ah Kim}
\affiliation{Department of Physics, Cornell University, Ithaca, New York 14853, USA}
\altaffiliation{Kavali Institute of Theoretical Physics}

\begin{abstract}
Recent experimental advances in using strain engineering to significantly alter the band structure of moderately correlated systems  offer opportunities and challenges to weak-coupling renormalization group (RG) analysis approaches for predicting superconducting instabilities. On one hand, the RG approach can provide theoretical guidance. On the other hand,  it is now imperative to better understand
how the predictions of the RG approach depends on microscopic and non-universal model details. Here we focus on the effect of band-selective mass-renormalization often observed in angle resolved photoemission spectroscopy. Focusing on a specific example of uniaxially strained $\rm{Sr_2RuO_4}$ we carry out the weak-coupling RG analysis from two sets of band structures as starting points: one is based on density functional theory (DFT) calculations and the other is based on angle-resolved photoemission spectroscopy (ARPES) measurements. 
Despite good agreement between the Fermi surfaces of the the two band structures 
we find the two sets of band structures to predict qualitatively different trends in the strain dependence of the superconducting transition temperature $T_c$ as well as the dominant channel.  
\end{abstract}

\maketitle

\section{Introduction}

Strain of magnitude that can significantly alter the band structure of correlated materials recently reached via bulk strain \cite{Hicks18042014,HicksVHS} or epitaxial strain\cite{bulatBRO} now offers a new axis of control beyond traditional means. 
This new dimension presents both opportunity and challenge for theory. 
On the bright side, weak-coupling renormalization group (RG) approaches\cite{Twostep,SriSRO,ScaffidiSROV,HicksVHS,IronScChubukov} that take the band structure based on microscopic information as starting points can now aspire to guide experimental efforts~\cite{BROsc}. Nonetheless, such proximity to reality puts higher bar on the theory. Especially, importance of better understanding
the sensitivity of the RG predictions against microscopic details cannot be understated. 

Here, we focus on the impact of band-specific mass renormalization that can affect the RG prediction for dominant pairing channel in a qualitative manner. It is well-known that band structures obtained using DFT inaccurately predict bandwidth. In a single-band system, this is often remedied through overall rescaling. Now growing interest in multi-band systems have presented a new challenge: the discrepancy in the band mass, often referred to as ``mass renormalization'', is often band selective\cite{bulatBRO}. Under such band-specific mass renormalization there is no simple way to reconcile the discrepancy between the dispersion of density functional theory (DFT) based band structure and that of angle-resolved photoemission spectroscopy (ARPES) measurements, even when the two band structures exhibit virtually identical Fermi surfaces (FSs). Now the critical question is the possible impact of such mass renormalization in possible superconducting instability. We investigate this issue in the context of strain-dependence of superconducting instability in $\rm{Sr_2RuO_4}$.

\begin{figure}[]
\subfigure[]{
	\includegraphics[width=5cm]{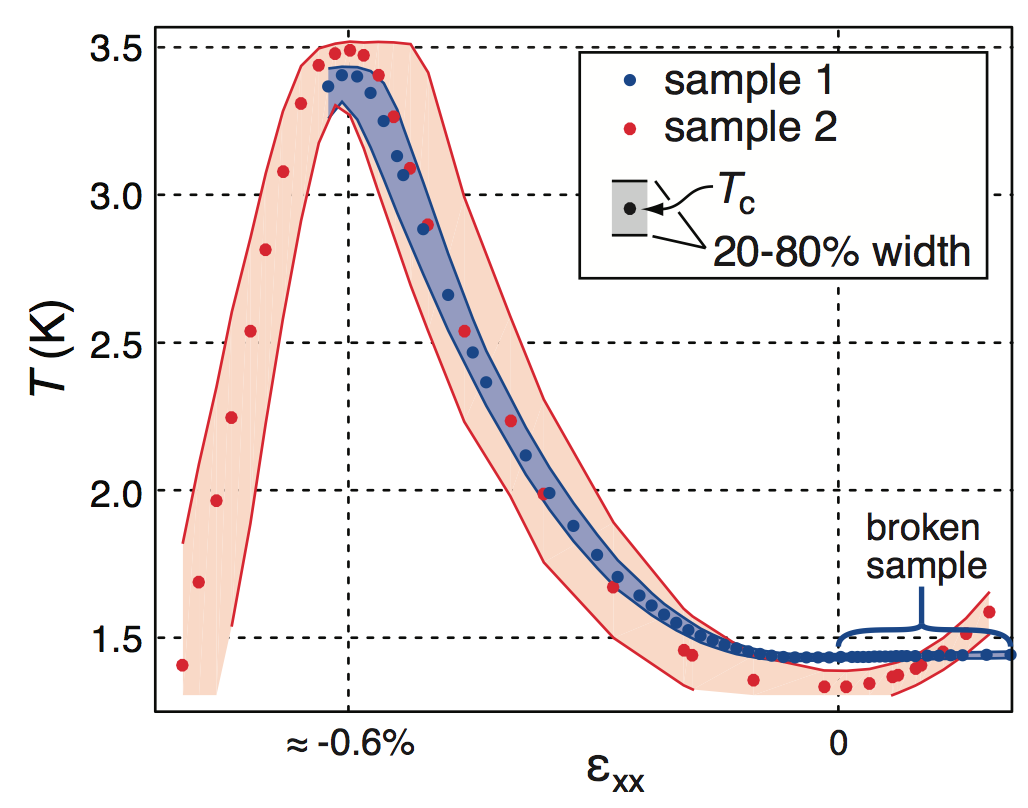}
	}
	\subfigure[]{
	\includegraphics[width=4cm]{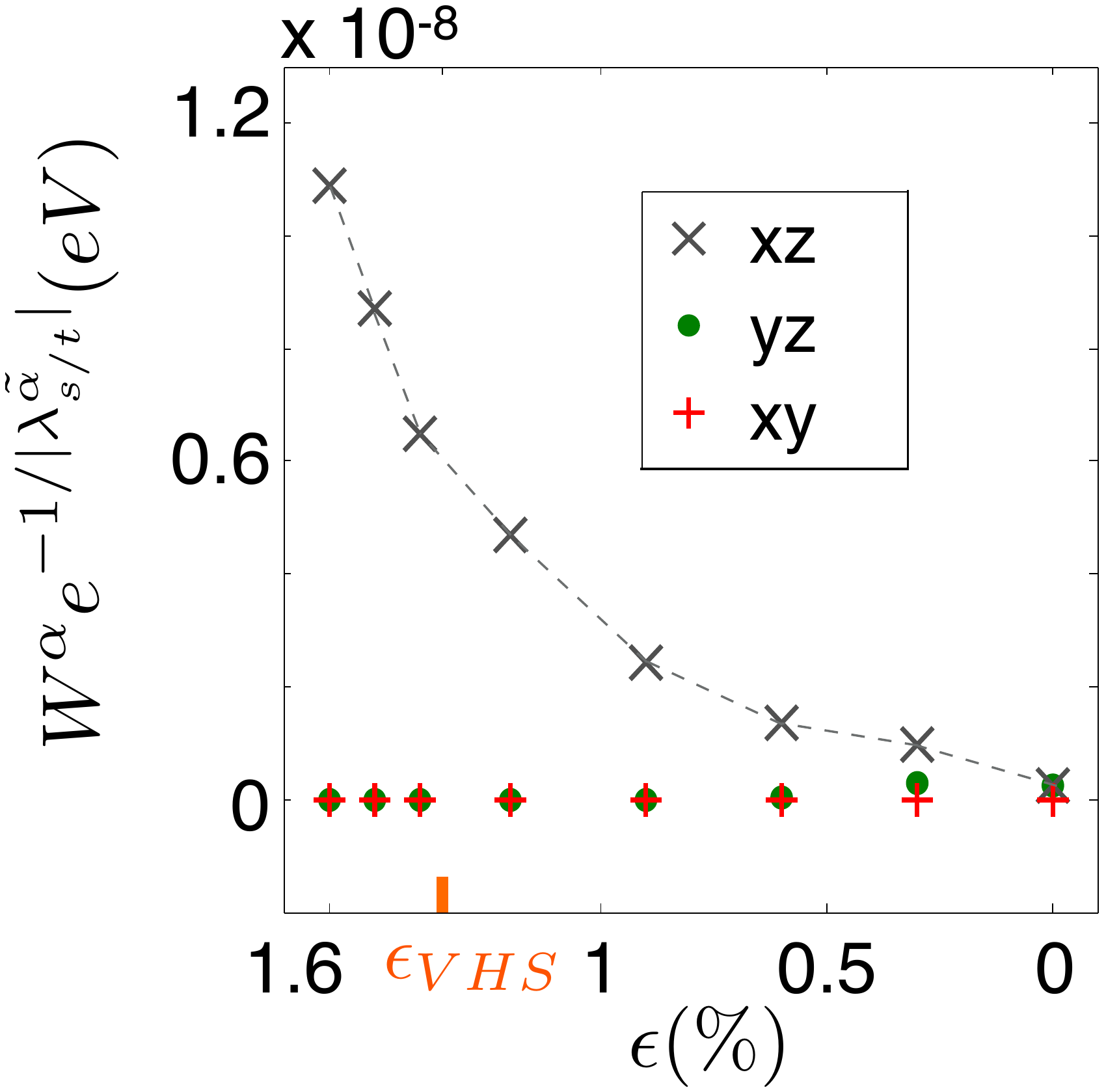}
	}
	\subfigure[]{
	\includegraphics[width=4cm]{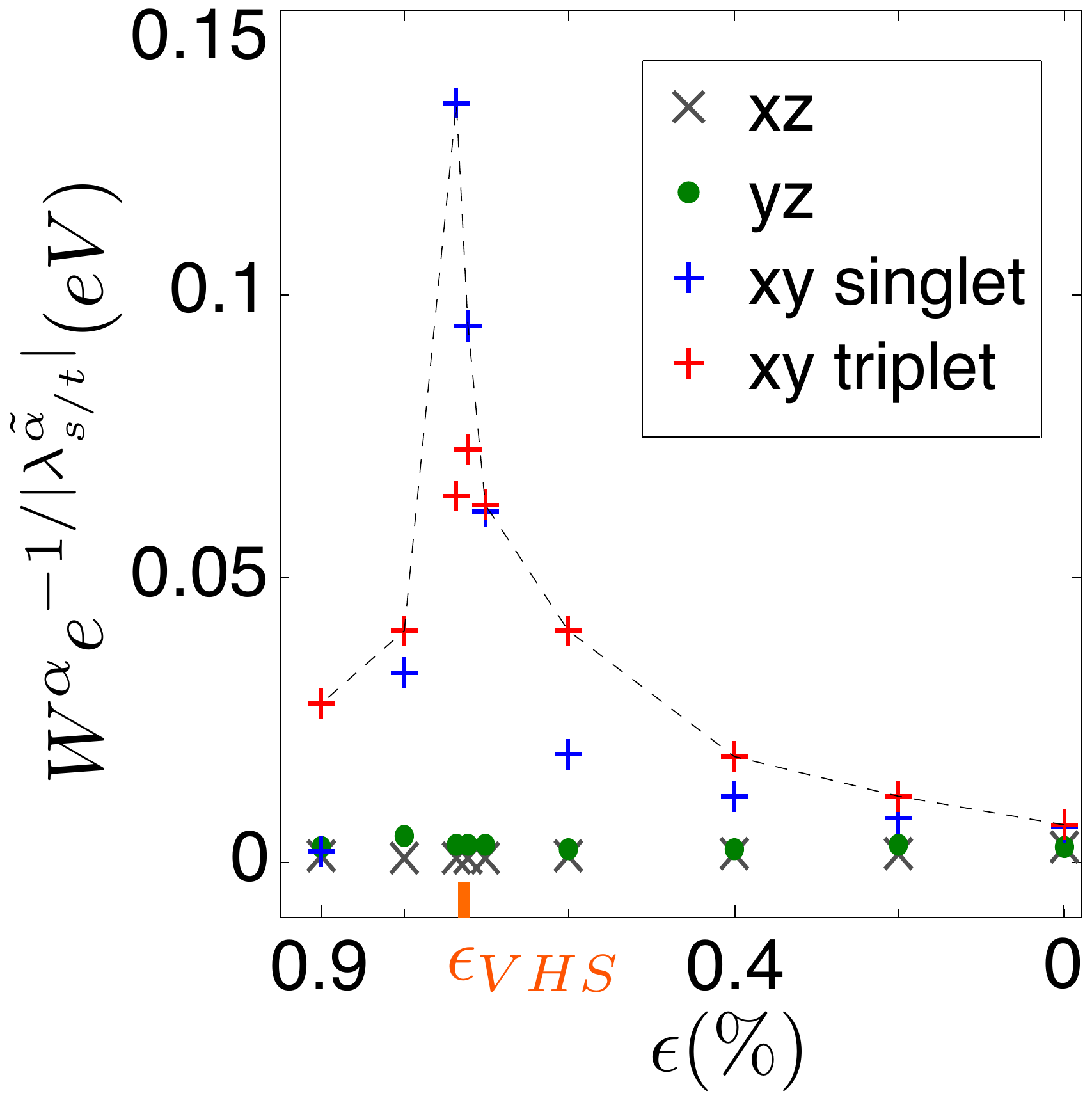}
	}
\caption{(a) The measured $T_c$ as a function of uniaxial strain presented in Ref.~\onlinecite{HicksVHS}.  (b-c) The estimate of $T_c\sim W^{\alpha}e^{-1/|\tilde{\lambda}^{\alpha}|}$ for different pairing channels as a function of uniaxial strain with $U=1$ eV using tight-binding parameters based on (b) the DFT results and (c) the ARPES data. Here, $\epsilon_{\rm{VHS}}$ is the strain amount at which the 2D band FS goes through Lifshitz transition, and the dashed line shows the expected transition temperature $T_c$. 
}
\label{Tc}
\end{figure}

Partially driven by the fact that $\rm{Sr_2RuO_4}$ is the leading candidate material for a two-dimensional (2D) topological superconductor\cite{SROreviewKallin,Ishida1998,Nelson12112004,Kidwingira24112006,PhysRevB.76.014526,PhysRevB.89.144504,Jang14012011}, recent strain-engineering efforts and careful study of experimental band structure focused on the material. In particular, \textcite{bulatBRO} reported band-specific mass renormalization. \textcite{bulatBRO} also reported that ruthenate films can undergo Lifshitz transition upon epitaxial biaxial strain of order 1.6\%\cite{bulatBRO,BROsc}. More recently, \textcite{HicksVHS} found the superconducting transition temperature ($T_c$) of $\rm{Sr_2RuO_4}$  to change upon a bulk compressive uniaxial strain  and peak at a certain magnitude [see Fig. \ref{Tc}(a)]. 
They further found that the upper critical field shows a decreased anisotropy at the $T_c$ maximum, which indicates the pairing to become spin-singlet at the point.   
Nonetheless, the large mass renormalization found in the ARPES data of biaxially strained ruthenates suggests possible discrepancies between the DFT-based band structures and actual band structure under uniaxially strain.
In particular, the band-selective mass renormalization in $\rm{Sr_2RuO_4}$ was found to be at the order of $30\%$ larger in the quasi-2D band than in the quasi-1D bands depend on the strain magnitude\cite{bulatBRO}.

The purpose of this article is to point out a fact that has been so far under-appreciated by the community: the perturbative RG results could be very sensitive to details in the input band structures. This would imply that extra attention is necessary in attempts to connect RG predictions with experiments. 
To demonstrate our point, we will take the uniaxially strained $\rm{Sr_2RuO_4}$ as an example to contrast the RG results based on the two sets of tight-binding parameters: one obtained from a DFT calculation, and the other from extrapolating the ARPES data in the absence of strain. 
Then, we will compare the two sets of results to the measured strain-dependent $T_c$\cite{HicksVHS}. 

\section{The model and approach}
\begin{figure}[!h]
\subfigure[]{
	\includegraphics[width=4cm]{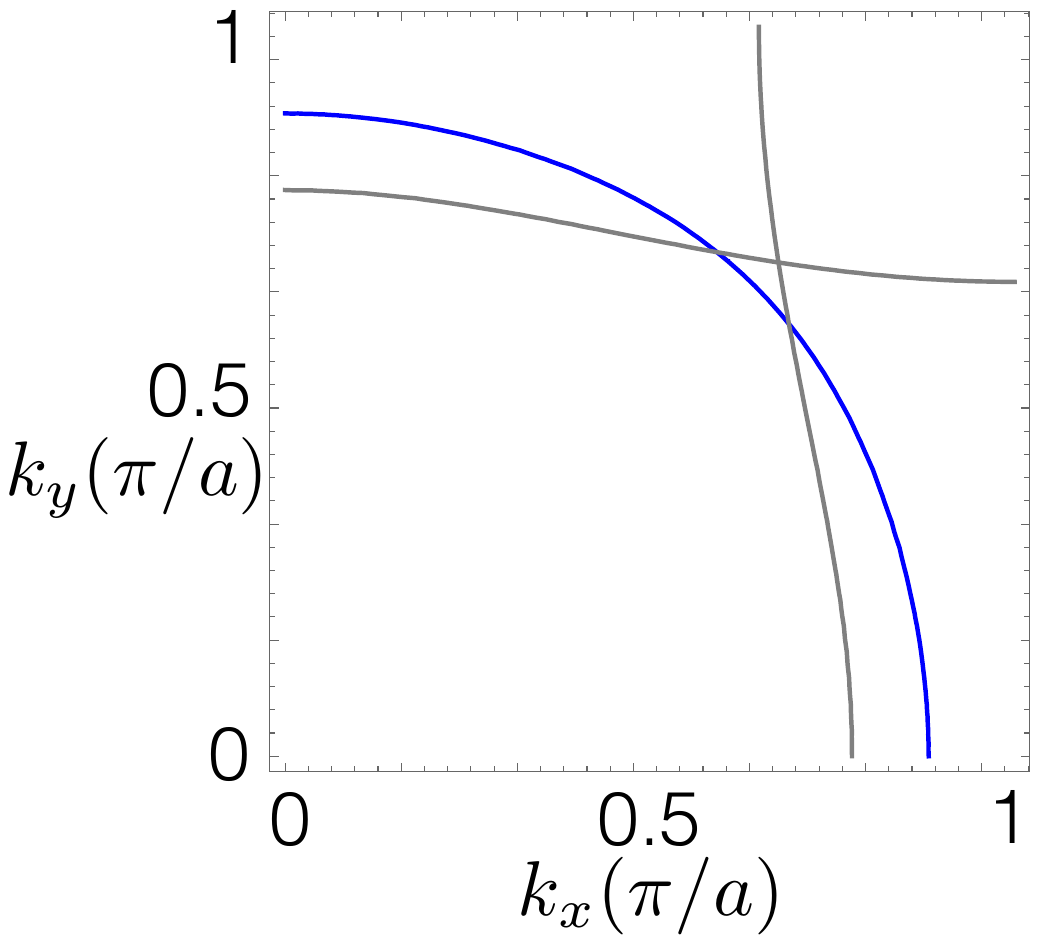}
	}
\subfigure[]{
	\includegraphics[width=4cm]{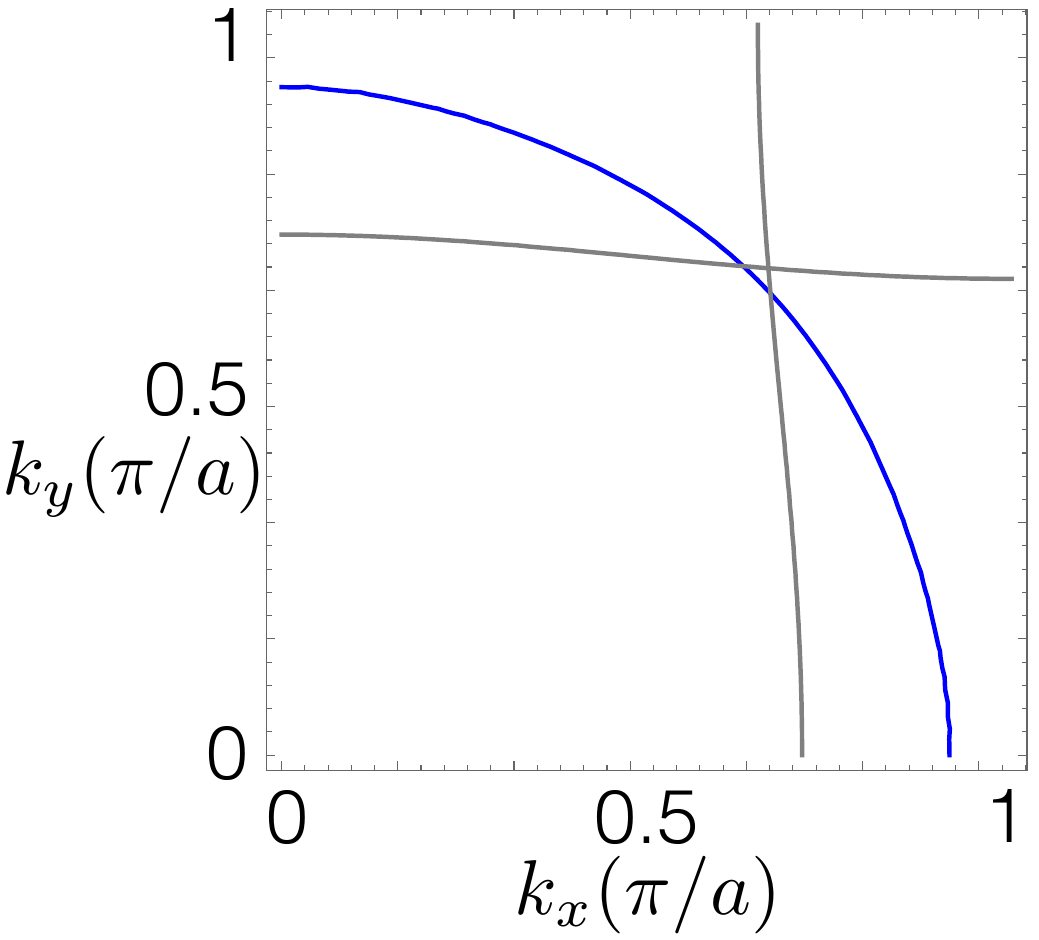}
	}
\caption{The FSs of unstrained $\rm{Sr_2RuO_4}$ obtained from the tight-binding model with parametrizations extracted from (a) DFT calculation (b) ARPES measurement. 
}
\label{FS}
\end{figure}

Our model for the uniaxially strained $\rm{Sr_2RuO_4}$ is a three-band Hubbard model derived from the Ru $t_{2g}$ orbitals $d_{xz}$, $d_{yz}$, and $d_{xy}$:
\begin{align}
H(\epsilon)= \sum_{\vec{k}\alpha\sigma}E^{\alpha}_{\vec{k}}(\epsilon)c^{\dagger}_{\vec{k},\alpha,\sigma}c_{\vec{k},\alpha,\sigma}
+U\sum_{i\alpha}n_{i,\alpha,\uparrow}n_{i,\alpha,\downarrow},
\label{eq:bareH}
\end{align} 
where $\epsilon<0$ denotes the compressive uniaxial strain along [100] direction. Here, $\vec{k}=(k_x,k_y)$, $\alpha=xz,yz,xy$, $\sigma=\uparrow,\downarrow$ denote the crystal momentum, the orbital index, and the spin respectively, and $n_{i,\alpha,\sigma}\equiv c^{\dagger}_{i,\alpha,\sigma}c_{i,\alpha,\sigma}$. 
We employ the following tight-binding parameterization for intra-orbital kinetic energies:
\begin{align}
&E^{xz}_{\vec{k}}(\epsilon)=-2t_x(\epsilon)\cos k_{x}-2t^{\perp}_y(\epsilon)\cos k_{y}-\mu_1(\epsilon)\nonumber\\
&E^{yz}_{\vec{k}}(\epsilon)=-2t_y(\epsilon)\cos k_{y}-2t^{\perp}_x(\epsilon)\cos k_{x}-\mu_1(\epsilon)\nonumber\\
&E^{xy}_{\vec{k}}(\epsilon)=-2t'_x(\epsilon)\cos k_x-2t'_y(\epsilon)\cos k_y\nonumber\\
&~~~~~~~~~~~~~-4t''(\epsilon)\cos k_x\cos k_y-\mu_2(\epsilon), 
\label{eq:E}
\end{align}
where we neglect the orbital-mixing terms \footnote{Although
\textcite{ScaffidiSROV} found the spin-orbit coupling to significantly affect the nature and mechanism of pairing in the unstrained system, the van Hove singularities which sit at point $X=(\pi,0)$ and $Y=(0,\pi)$ lie in the region of the FS where orbital characters are well-defined\cite{SOCMackenzie,SROreviewKallin}. 
Hence, we expect the absence of orbital-mixing terms in our model would not affect our conclusions in a qualitative manner.}. 
The dispersions of the three bands in Eq.~\eqref{eq:E} lead to two quasi-1D FSs comprising the Ru orbitals $d_{xz}$ and $d_{yz}$, and
one quasi-2D FS comprising the 
Ru orbital $d_{xy}$. 
For the bare interaction, we focus on the 
repulsive intra-orbital on-site repulsion $U>0$\cite{SriSRO} given the experimentally observed unconventional pairing in as-grown $\rm{Sr_2RuO_4}$ \cite{Ishida1998,Nelson12112004,Kidwingira24112006}. 

Our model above differs from the model in Ref.~\onlinecite{HicksVHS} in that 
the latter has emphasized the inter-band coupling. Furthermore, starting from their DFT-based band structure, they found the quasi-1D bands to be the leading pairing channels although it is the 2D band that goes through the Lifshitz transition. Therefore, the inter-band repulsion $U'$ of a significant magnitude  of $0.84U$ was cruicial for the predominantly 1D-band-driven superconductivity to nevertheless show the experimentally observed peak in $T_c$ as a function of uniaxial strain while the 2D band is closer to the van Hove singularity in the model of Ref.~\onlinecite{HicksVHS}. Here, we focus on the results in the absence of 
the inter-orbital repulsion $U'$. Nevertheless, we have checked that the inter-orbital $U'\leq 0.5U$ makes no qualitative difference to the results we report in this paper.

Our main concern is the effects on the pairing instability from the mass renormalization, which is often present in measured band structures, despite the calculated and measured FSs could be qualitatively similar [see Fig.~\ref{FS}]. Thus in the following, 
we contrast and compare the RG predictions starting from two sets of tight-binding parameters $E^{\alpha}_{\vec{k}}(\epsilon)$ in Eq.~\eqref{eq:E}: 1) the parameters fitted to DFT calculations with varying degree of strain, and 2) the parameters fitted to the available unstrained ARPES data and strained appropriately. For the first set of parameters we performed DFT calculations by fixing the [100] lattice constants to the desired strain value and by letting all internal parameters as well as transverse lattice constant to fully relax. All our DFT calculations were performed with VASP \cite{DFT1,DFT2}, using the PBEsol Exchange-correlation functional, a plane-wave basis cutoff of 520 eV and a 12x12x12 k-point sampling of the Brillouin zone. The band structure thus obtained was then used to fit the tight-binding model in Eq.~\eqref{eq:E}.

For the second set of parameters, 
we use the parameters extracted from the ARPES data of an unstrained $\rm{Sr_2RuO_4}$\cite{bulatBRO} at zero strain $\epsilon=0$. As no ARPES data is currently available under uniaxial strain $\epsilon< 0$, we determine the tight-binding parameters under strain by extrapolating the unstrained ARPES-extracted parameters. For this, we determine the percentage change of each parameter $p(\epsilon)$ under strain from the first set of DFT-extracted parameters by $p(\epsilon)\equiv t_x^{DFT}(\epsilon)/t_x^{DFT}(0) -1$. We then estimate each strained parameter starting from the ARPES-measured parameter for unstrained system as $t_x(\epsilon)=t_x(0)[1+p(\epsilon)]$. 
The key difference between the first and the second set of parameters is the band-selective mass renormalization that has been measured in the ARPES data of Ref.~\onlinecite{bulatBRO} in the absence of strain. 
Although it is well-known that DFT often underestimates band mass, 
band-selective mass renormalization in multi-band system poses challenges that have been under-appreciated. 
It turns out that the mass renormalization is focused on 2D bands in $\rm{Sr_2RuO_4}$, which enhances the density of states of the 2D band at the Fermi level substantially.\cite{bulatBRO,BROsc}\footnote{
Here for the second set of parameters, we have effectively assumed the bias in mass renormalization stays constant as the uniaxial strain increases though the ARPES data is yet not available.}

To study the dominant pairing channels under strain, we then carry out a two-step perturbative RG analysis with the microscopic model being Eq.~\eqref{eq:bareH} and \eqref{eq:E} with the two sets of tight-binding parameters.  
For completeness, we now briefly review the perturbative two-step RG approach\cite{Twostep,SriSRO} we adopt.
In the first step, we numerically evaluate the effective pairing vertices in different channels at some intermediate energy scale $E=\Lambda_0$ near the Fermi level by integrating out the higher-energy modes down to $\Lambda_0$. 
Up to the one-loop order, the singlet and triplet effective pairing vertices $\Gamma_{s/t}^{\alpha}(\hat{k},\hat{k'})$ at energy $\Lambda_0$ are related to the repulsive bare interaction $U$ and the non-interacting static particle-hole susceptibilities $\Pi_{ph}^{\alpha}(\vec{q})$ for band $\alpha$ at momentum $\vec{q}$ through
\begin{align}
\Gamma_s^{\alpha}(\hat{k},\hat{k'})=&U+U^2\Pi_{ph}^{\alpha}(\hat{k}+\hat{k'}),
\label{eq:Gammas}
\end{align}
and 
\begin{align}
\Gamma_t^{\alpha}(\hat{k},\hat{k'})=&-U^2\Pi_{ph}^{\alpha}(\hat{k}-\hat{k'}), 
\label{eq:Gammat}
\end{align}
where $\hat{k}^{(')}$ are the outgoing (incoming) momenta on the FS of band $\alpha$.
Now, the pairing tendency of band $\alpha$ in the singlet and triplet channels can be quantified by the most negative eigenvalue $\tilde{\lambda}_{s/t}^{\alpha}\equiv\lambda_{s/t}^{\alpha}(E=\Lambda_0)$ of a dimensionless matrix $g_{s/t}^{\alpha}(\hat{k},\hat{k'})$, which is a
 product of the density of states $\rho^{\alpha}$
 on the Fermi surface of the band $\alpha$
and the normalized effective pairing vertices at the energy scale $\Lambda_0$:
\begin{align}
g_{s/t}^{\alpha}(\hat{k},\hat{k'})=\rho^{\alpha}\sqrt{\frac{\bar{v_F}^{\alpha}}{v_F^{\alpha}(\hat{k})}}\Gamma_{s/t}^{\alpha}(\hat{k},\hat{k'})\sqrt{\frac{\bar{v_F}^{\alpha}}{v_F^{\alpha}(\hat{k'})}}.
\label{eq:gmatrix}
\end{align}
Here,  $v_F^{\alpha}(\hat{k})$ is the magnitude of Fermi velocity at $\hat{k}$, and $\frac{1}{\bar{v_F}^{\alpha}}\equiv\int\frac{d\hat{p}}{S_f^{\alpha}}\frac{1}{v_F^{\alpha}(\hat{p})}$ with $S_F^{\alpha}\equiv\int d\hat{p}$ being the FS `area' of band $\alpha$.
In the second step, we study the evolution of the most negative eigenvalues $\lambda_{s/t}^{\alpha}(E)$ for different channels ($\alpha$, $s/t$) as the energy $E$ lowers from $\Lambda_0$ to $0$. Given the well-known RG flow for the Cooper instability, 
 $\frac{d\lambda_{s/t}^{\alpha}}{dy}=-(\lambda_{s/t}^{\alpha})^2$ with the RG running parameter being  $y\equiv\log(\Lambda_0/E)$\cite{ShankarRG}, 
we can relate $T_c$ to the critical energy scale at which the most divergent $\lambda_{s/t}^{\alpha}(y)$ among all channels diverges as\cite{SriSRO}
\begin{align}
T_c\sim W^{\alpha}e^{-1/|\tilde{\lambda}|},
\label{eq:Tc}
\end{align}
where $W^{\alpha}$ is the bandwidth of the dominant band $\alpha$, and $\tilde{\lambda}$ is the most negative $\tilde{\lambda}_{s/t}^{\alpha}$ among all channels.

\section{Results}
Using the first set of tight-binding parameters obtained from DFT, we find the critical energy scale defined in Eq.~\eqref{eq:Tc} to increase \textit{monotonically} with the compressive 
uniaxial strain in [100] direction $\epsilon<0$ [see Fig.~\ref{Tc}(b)].
This is because the active band is the $yz$-orbital-based 1D band whose density of states monotonically increase with the compressive strain, as opposed to that of the $xy$-orbital-based 2D band which peaks at the strain amount $\epsilon_{\rm{VHS}}$, where the 2D band FS goes through Lifshitz transition. 
The 1D band dominates over the 2D band despite the fact that the 2D band density of states $\rho^{xy}$ is slightly larger than that of the 1D band $\rho^{yz}$. This is  
because the particle-hole susceptibility of the 1D band peaks sharply at $\vec{q}\sim(\pi,2k_F)$ due to the high degree of nesting.  This is a feature shared between our DFT-based band structure and the DFT-based band structure used by \textcite{HicksVHS}. Similarly,  \textcite{HicksVHS} also found the 1D band 
to dominate the pairing instability. It was only through a substantial inter-orbital coupling $U'$ could \textcite{HicksVHS} find the $T_c$ scaling to peak riding on the van Hove singularity touched by the 2D band FS. 

By contrast, the $T_c$ calculated using the second set of parameters based on ARPES data peaks as a function of strain even in the absence of any inter-orbital coupling. This is because the 2D band is now the active band due to the mass renormalization which is substantially more severe on the 2D band\cite{bulatBRO}. Hence, when the 2D FS goes through Lifshitz transition at the strain amount $\epsilon_{\rm{VHS}}$, $T_c$ peaks [see the dashed line in Fig.~\ref{Tc}(c)]. 
Note that in the close vicinity of the van Hove singularity at $(\pm\pi,0)$, the parity-even singlet dominates over the parity-odd triplet pairing tendency as the latter is expected to be suppressed by the symmetry\cite{YaoVHSII,ChubukovHexSC}\footnote{The triplet tendency is expected to vanish right at $\epsilon_{\rm{VHS}}$ which is not captured in Fig.~\ref{Tc}(c) because the perturbative RG analysis cannot access the non-perturbative regime.}.
Interestingly, the prediction for the peak in $T_c$ and the dominance of the singlet pairing in the close vicinity of the peak agrees with what was observed in the experiment [see Fig.~\ref{Tc} (a)]. The fact that key experimental features are robustly reproduced by the RG prediction with a simple model for interactions is rather appealing. 

To summarize, we investigated how perturbative RG predictions for superconducting instability depends often on the understated aspects of the band structure beyond Fermi surface. We found that in a multi-band model, balance between mass renormalizations of different bands can change the balance between different pairing channels, and thus the qualitative trends of pairing properties under external knobs.  
Motivated by recent experimental findings (1) $T_c$ peaking at a finite percentage of uniaxial strain, and (2) singlet pairing near the peak, we investigated the specific example of uniaxially strained $\rm{Sr_2RuO_4}$ with two sets of band structures. We found the two band structures to host qualitatively different  trends in $T_c$ as a function of strain: while the DFT-based band structure fails to reproduce the observed peak in $T_c$ in the absence of strong inter-orbital interaction\cite{HicksVHS},  the ARPES-based band structure reproduces the observed peak even with a simple Hubbard type model.  This shows band-selective mass renormalizations can affect balance between different superconducting channels, hence calls for realistic band-structure information to accompany strain engineering studies.\\

{\it Acknowledgement --} 
Y-TH was supported by the Cornell Center for Materials Research with funding from the NSF MRSEC program (DMR-1120296) and E-AK was supported by U.S. Department of Energy, Office of Basic Energy Sciences, Division of Materials Science and Engineering under Award DE-SC0010313. E.-A.K. acknowledges Simons Fellow in Theoretical Physics Award \#392182 and thanks hospitality of KITP supported by Grant No. NSF PHY11- 25915. AFR and CJF acknowledge support from the NSF grant no. DMR-1056441.

%

\end{document}